\documentclass[reprint,superscriptaddress,secnumarabic,aps,prd,longbibliography]{revtex4-2}

\usepackage{graphicx}
\usepackage{dcolumn}
\usepackage{bm}
\usepackage[mathlines]{lineno}
\usepackage{amsmath}
\usepackage[%
colorlinks=true,
urlcolor=blue,
linkcolor=blue,
citecolor=blue
]{hyperref}
\usepackage{array}
\usepackage{booktabs}
\usepackage[utf8]{inputenc}
\usepackage[T1]{fontenc}
\usepackage{array,float}
\usepackage{mathptmx}
\usepackage{amssymb}
\usepackage{eurosym}
\usepackage{mathrsfs}
\usepackage{tikz}
\usepackage{siunitx}
\usepackage{textcomp}
\usepackage[caption=false]{subfig}

\newcolumntype{C}[1]{>{\centering\arraybackslash}m{#1}}

\newcommand{\tcmp}{$T_{\text{comp}}$}

\begin{document}

\title[ Magnetic order and magneto-transport in half-metallic ferrimagnetic Mn$_y$Ru$_x$Ga thin films ]{Magnetic order and magneto-transport in half-metallic ferrimagnetic Mn$_y$Ru$_x$Ga thin films}

\author{K.E. Siewierska}
\email{siewierk@tcd.ie}
\author{G. Atcheson} 
\author{A. Jha}
\affiliation{ 
	School of Physics and CRANN, Trinity College Dublin, Ireland
}%
\author{K. Esien} 
\affiliation{ 
	Queen’s University Belfast, Northern Ireland, United Kingdom
}%
\author{R. Smith} 
\author{S. Lenne} 
\author{N. Teichert}
\author{J. O'Brien}
\author{J.M.D. Coey}
\author{P. Stamenov} 
\author{K. Rode}
\affiliation{ 
	School of Physics and CRANN, Trinity College Dublin, Ireland
}%


\begin{abstract}
\noindent The ruthenium content of half-metallic Mn$_2$Ru$_x$Ga thin films, with a biaxially-strained inverse Heusler structure, controls the ferrimagnetism that determines their magnetic and electronic properties. An extensive study of Mn$_y$Ru$_x$Ga films on MgO (100) substrates with $1.8 \leq y \leq 2.6$ and $x = 0.5$, 0.7 or 0.9, including crystallographic, magnetic order, magneto-transport and spin polarisation is undertaken to map specific composition-dependent properties in this versatile ternary system. A comparison of experimental densities obtained from X-ray reflectivity with calculated densities indicates full site occupancy for all compositions, which implies chemical disorder. All moments lie on the Slater-Pauling plot with slope 1 and all except $x = 0.5$, $y = 2.2$ exhibit magnetic compensation at \tcmp~below  500~K. The coercivity near \tcmp~exceeds 10~T. Increasing the Mn or Ru content raises \tcmp, but increasing Ru also decreases the spin polarisation determined by point contact Andreev reflection.  Molecular field theory is used to model the temperature dependence of the net ferrimagnetic moment and three principal exchange coefficients are deduced. Marked differences in the shape of anomalous Hall and net magnetisation hysteresis loops are explained by substantial canting of the small net moment by up to \SI{40}{\degree} relative to the $c$-axis in zero field, which is a result of slight non-collinearity of the Mn$^{4c}$ sublattice moments due to competing intra-sublattice exchange interactions arising from antisite disorder and excess Mn in the unit cell. Consequences are reduced spin polarisation and an enhanced intrinsic contribution to the anomalous Hall effect. The systematic investigation of the physical properties as a function of $x$ and $y$ will guide the selection of compositions to meet the requirements for magnonic and spintronic MRG-based devices.
\end{abstract}

\maketitle

\section{\label{sec:level1}Introduction}

\begin{figure}
	\centering
	\includegraphics[scale=0.2]{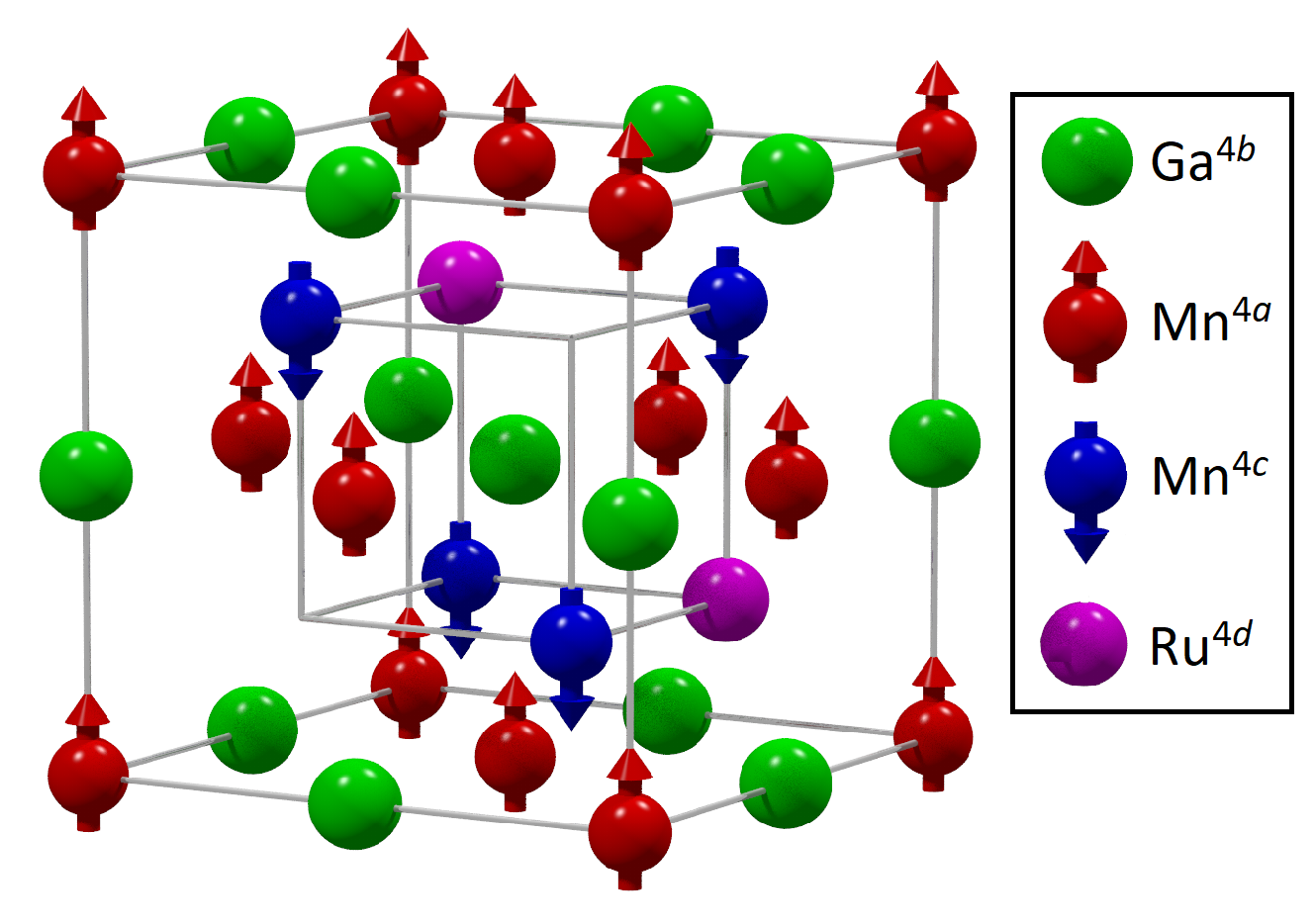}
	\caption{Model of the inverse Heuser \textit{XA} structure of Mn$_2$Ru$_{0.5}$Ga.}
	\label{figure:model}
\end{figure}

Zero-moment ferrimagnetic half-metals (ZMHM) are attractive materials for applications in spintronic devices. \cite{Betto.2016,Finley2020} They offer advantages over their antiferromagnetic or half-metallic ferromagnetic counterparts. \cite{GRAF20111} After their theoretical prediction in 1995\cite{vanLeuken.1995}, various  candidate ZMHM materials were proposed over the years based on DFT calculations, including C1$_b$ CrMnSb, L2$_1$ Fe$_2$VGa and D0$_3$ Mn$_3$Ga \cite{Wurmehl, Hu2012, Galanakis}. Unfortunately, the predicted alloys either crystallised in a different structure with no spin gap like Mn$_3$Ga \cite{Kharel}, decomposed into a mixture of simpler phases like CrMnSb or were found be either nonmagnetic or very weak feromagnets, like Fe$_2$VGa \cite{Hakimi}. It seemed there might be some reason why the ZMHM was so elusive \cite{Hu2012}, The  first material to show evidence of the long-sought magnetic behaviour was Mn$_2$Ru$_{0.5}$Ga.\cite{Kurt.2014} Extrapolating between a half-Heusler Mn$_2$Ga and a full-Heusler Mn$_2$RuGa,  Mn$_2$Ru$_{0.5}$Ga was thought to have 21 valence electrons with  vacancies on half the  Ru$^{4d}$ sites, as shown in Fig. \ref{figure:model}. \cite{Kurt.2014} Since its discovery, a few other members of ZMHM material class have been demonstrated, including thin film Mn$_3$Al  \cite{Jamer.2017} and bulk Mn$_{1.5}$V$_{0.5}$FeAl \cite{Stinshoff2017}. 
The Mn$_2$Ru$_x$Ga (MRG) thin films grown on MgO (001) substrates, where the cubic structure of the inverse Heusler alloy is distorted by biaxial strain, offer perpendicular magnetic anisotropy (PMA) that allows efficient detection and manipulation of the magnetic state. Cubic MRG would crystallise in the \textit{F}$\bar{4}$3m (216) space group, with 4 formula units (f.u.) per unit cell, which is adopted when referring to crystallographic vectors and positions, but the symmetry is reduced to \textit{I}$\bar{4}$m2 (119) by the tetragonal distortion. In the ideal inverse \textit{XA} Heusler structure, $4a$ and $4b$ sites are occupied by Mn and Ga, respectively, whereas $4c$ and $4d$ sites are fully occupied by Mn and Ru. \cite{Kurt.2014} Transport and magneto-optic properties are dominated by the Mn$^{4c}$ sublattice because there are few, if any, Mn$^{4a}$ states at the Fermi level. \cite{Kurt.2014,Betto.2016,Thiyagarajah.2015} The two manganese sublattice magnetic moments cancel at the compensation temperature \tcmp. Nevertheless, magnetic domains can be observed by the magneto-optic Kerr effect (MOKE) even at temperatures where there is little or no net magnetisation. \cite{Siewierska_MOKE,Teichert2020} As the net moment tends to zero, the anisotropy field becomes large, resulting in high-frequency spin dynamics with low Gilbert damping. \cite{Kurt.2014,Betto.2016} A maximum zero-field resonance frequency of $\approx 160 $~GHz and a Gilbert damping constant $\alpha \approx 0.02$ have been reported. \cite{Betto.2016,Bonfiglio2019}

Remarkably efficient charge/spin conversion and large spin-orbit fields per current density $\frac{\mu_0 \bm{H}_{\text{eff}}}{\bm{j}}$ related to spin currents have recently been measured in single thin films of MRG, where $\frac{\mu_0 \bm{H}_{\text{eff}}}{\bm{j}}$ approaches $0.1 \times 10^{-10}$ T/Am$^{-2}$ in the low-current density limit. This is almost a thousand times the \O{}rsted field and one to two orders of magnitude greater than $\frac{\mu_0 \bm{H}_{\text{eff}}}{\bm{j}}$ in heavy metal/ferromagnet bilayers. \cite{lenne2019giant} The efficiency of current induced spin-orbit torque switching in Ru/MRG/MgO has been found to be similar to that of a ferromagnet, e.g. Ta/CoFeB/MgO \cite{Cubukcu2014} or Pt/Co/AlO$_x$ \cite{Liu2012}, despite larger MRG film thickness and coercivity \cite{Finley2019}. Furthermore, thermal single-pulse all-optical toggle switching at ultra short timescales ($<10$~ps) has been recently demonstrated in MRG, making it the first non-Gd based material to exhibit this effect. \cite{banerjee2019single,Davies2020,banerjee2021} MRG-based devices could offer a practical solution to the current problem of chip-to-chip generation and detection of electromagnetic radiation in the 0.1 THz - 10 THz frequency range, known as the `THz gap', provided sufficiently high magnetoresistive effects can be achieved. \cite{Betto.2016,Borisov.2016} 

Each spintronic application requires a specific set of material properties. For example, efficient single-pulse all optical toggle switching (SPAOS) at room temperature (RT) in MRG films requires \tcmp~to be just above RT. \cite{banerjee2019single} High spin polarisation and perfectly crystalline films with smooth surfaces are desirable for MRG-based magnetic tunnel junctions. Earlier studies have investigated the variation of some of these properties with $x$ in Mn$_2$Ru$_x$Ga. \cite{Kurt.2014,Zic.2016,Thiyagarajah.2015,Betto.2015} The results show that an increase in Ru content increases tetragonal distortion of the cubic unit cell on MgO and improves the wetting of the film. These qualities are useful for all applications, and particularly for the fabrication of nanostructured devices. \cite{Zic.2016,Thiyagarajah.2015} 

The goal of the present work is to help identify compositions which exhibit the best combinations of properties for specific applications. Thin thin films are prepared with varying Ru and Mn content, while keeping the Ga content fixed. The dominant contribution of Mn$^{4c}$ sites to the density of states at the Fermi level allows us to disentangle net and sublattice magnetisations with the aim of establishing non-collinearity of the Mn$^{4c}$ sublattice moments, which results in the canting of net moment.

\section{\label{sec:level2}Methodology}

Epitaxial thin films of Mn$_y$Ru$_x$Ga (MRG) with $x + y \approx 3$ were grown by DC magnetron sputtering on $10 \times 10 $~mm$^2$ (100) MgO substrates in our Shamrock sputtering system \cite{Kurt.2014,Betto.2016}. The base pressure of the system was 10$^{-8}$~Torr. Films were co-sputtered in argon onto single-side polished substrates maintained at \SI{350}{\degreeCelsius} from three 75~mm targets of Mn$_2$Ga, Ru and either MnGa or Mn$_3$Ga. Deposition rates from each target were calibrated and used to determine the values of $x$ and $y$ in the formula. Films were capped in-situ with a 3~nm layer of AlO$_x$ deposited at room temperature to prevent further oxidation. Film thickness and rms roughness were determined by low- angle X-ray scattering in a Panalytical X'Pert Pro diffractometer, and thickness was found by fitting the interference pattern using X'Pert Reflectivity Software. A Bruker D8 Discovery X-ray diffractometer with a copper tube (K$_{\alpha}$ wavelength~$ = 154.06$~pm) and a double-bounce Ge [220] monochromator on the primary beam was used to determine the diffraction patterns of the thin films. Reciprocal space maps were obtained on the same diffractometer around the (113) MgO reflection. 

Magnetisation measurements with the field applied perpendicular or parallel to the surface of films mounted in a straw were made using a 5~T Quantum Design SQUID magnetometer. Data included hysteresis loops and thermal scans from 10~K - 400~K. They were corrected for the magnetism of the substrate. Thermal scans in zero field after saturation of the magnetisation at room temperature were used to determine the compensation temperatures of the films. For Kerr imaging, an Evico Magnetics wide-field Kerr microscope with a 10x/0.25 objective lens was used. All loops were measured with polar sensitivity and red LED light with a field applied out-of-plane. Faraday rotation was compensated during the measurement using a feature which readjusts the analyser position relative to a mirror reference.\cite{Soldatov2017} Samples were heated from 300~K - 500~K on a temperature-controlled microscope stage.

Electrical measurements were made in a 1~T GMW electromagnet under ambient conditions. Silver wires were cold-welded to the films with indium and the current used was 5~mA. High field data were obtained in a 14~T Quantum Design Physical Property Measurement System (PPMS\textsuperscript{TM}). The films there were contacted with silver paint. The 4-point Van der Pauw geometry was used to determine both the Hall resistivity and the longitudinal resistivity of the films. 

Point contact Andreev reflection (PCAR) measurements were made in the PPMS using a mechanically sharpened Nb tip. Landing the tip onto the sample surface is controlled by an automated vertical Attocube piezo-stepper. Two horizontal Attocube\textsuperscript{TM} steppers are used to move the sample laterally to probe a pristine area. The differential conductance spectra were fitted using a modified Blonder-Tinkham-Klapwijk model, as detailed elsewhere.\cite{PCAR_1,PCAR_2}

\section{\label{sec:level3}Results}

\begin{figure}[h]
	\centering
	\subfloat{\label{XRR}\includegraphics[scale=0.22]{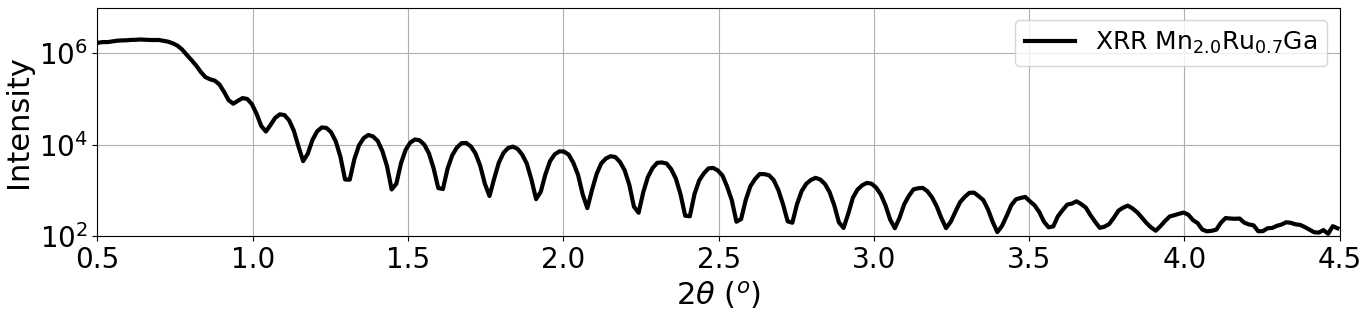}}\\
	\subfloat{\label{XRD}\includegraphics[scale=0.22]{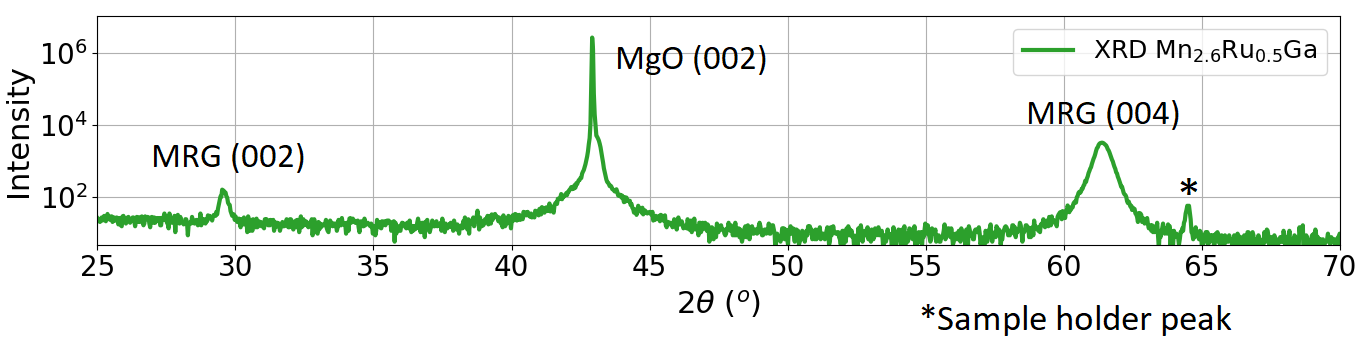}}
	\caption{X-ray analysis of MRG thin films: (a) XRR pattern of Mn$_{2.0}$Ru$_{0.7}$Ga and (b) XRD pattern of Mn$_{2.6}$Ru$_{0.5}$Ga. The peak marked * is due to the sample holder.} 
\end{figure}

Typical small-angle X-ray scattering data shown in Fig. \ref{XRR} attest to the film quality. Film and cap layer thickness values for all films were $50 \pm 5$~nm and $3 \pm 2$~nm, respectively and root mean square roughness were deduced from these data. The films are tetragonally elongated by about 1~\% as a result of biaxial compressive strain imposed by the substrate. A typical X-ray diffraction pattern in Fig. \ref{XRD} shows only the $c$-axis (002) and (004) reflections as expected from a perfectly textured film. The growth axis corresponds to the crystallographic $c$-axis. Broadening of the reflections is mainly due to the film thickness.

\begin{figure}[h]
	\centering\includegraphics[scale=0.28]{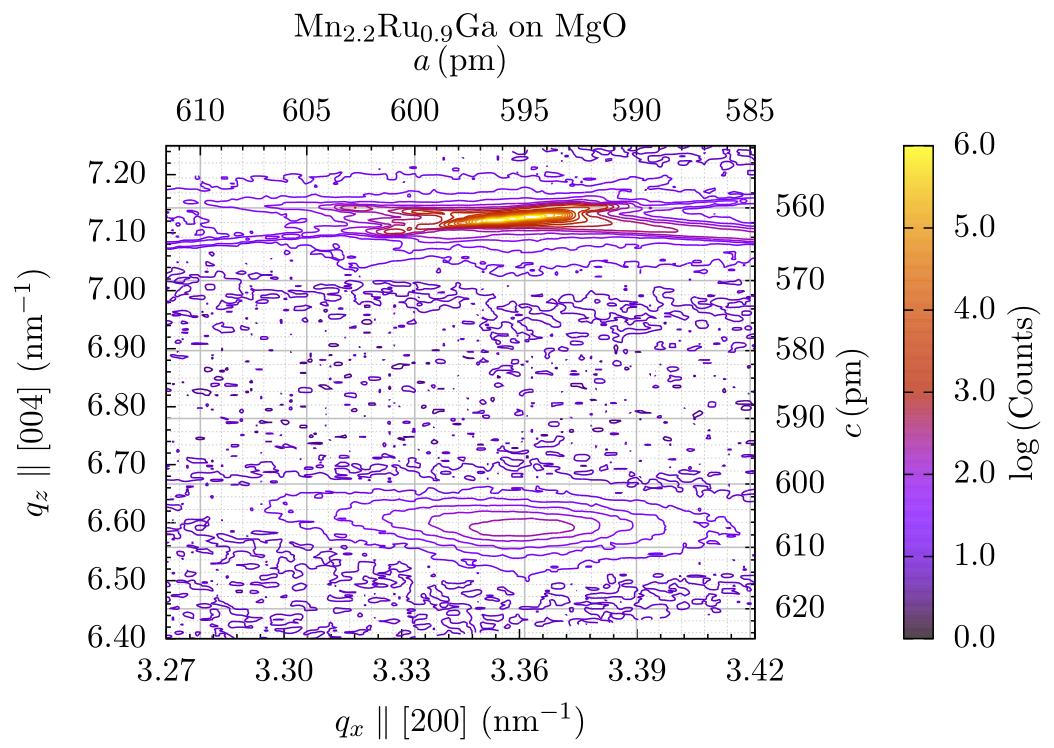}
	\caption{Reciprocal space map of Mn$_{2.2}$Ru$_{0.9}$Ga of MgO (113) and MRG (204) reflections. Lattice parameters are calculated with respect to the MRG unit cell. } 
	\label{RSM}
\end{figure} 

In order to determine both unit cell parameters, and visualise the strain in the film, reciprocal space maps were recorded around the [113] MgO reflection. An example is shown in Fig. \ref{RSM}. The $a$-parameter of MRG is constrained by the substrate to be $\sqrt{2}a_{MgO} = 595$~pm, whereas the c-parameter expands by about 1~\%.

\begin{figure}[h]
	\centering\includegraphics[scale=0.23]{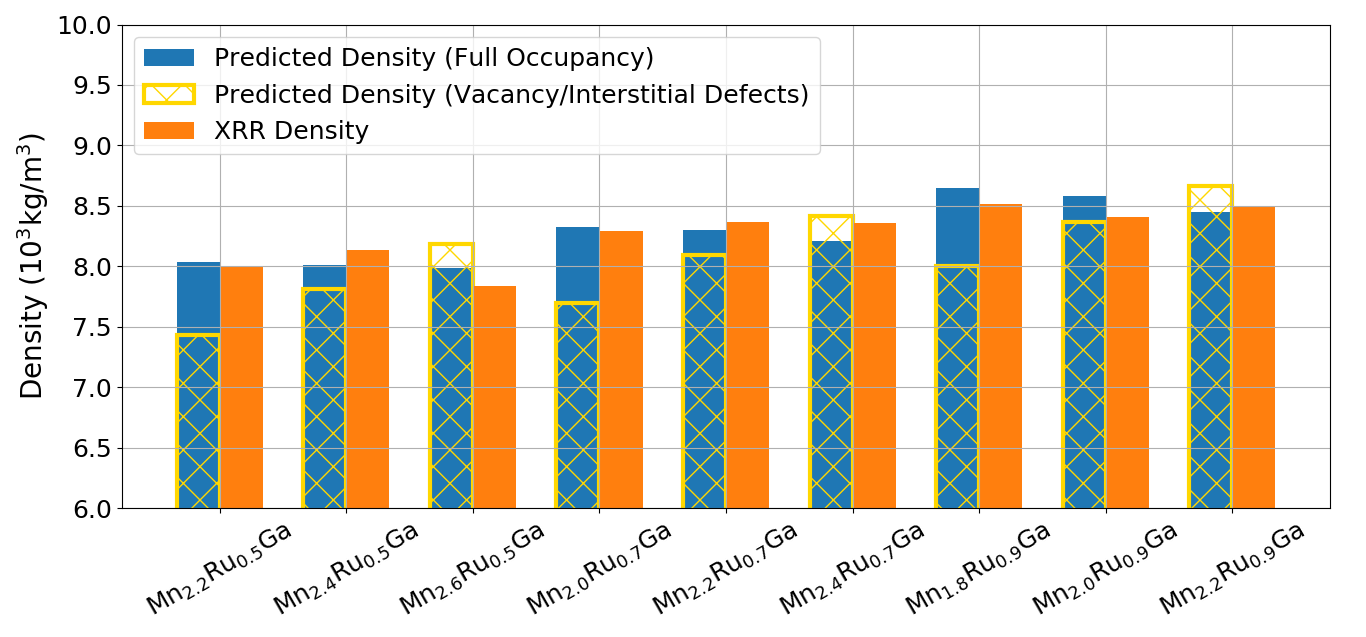}
	\caption{Calculated (blue) and experimental (orange) densities for the nine MRG films. The yellow crossed bars show densities predicted if there were vacancies/interstitials in the structure. } 
	\label{Density}
\end{figure}

\begin{figure}[h]
	\centering\includegraphics[scale=0.34]{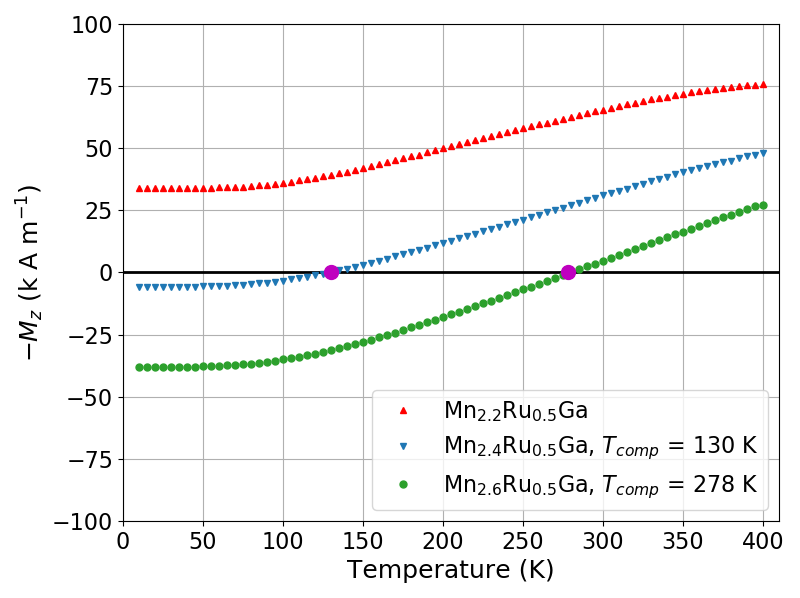}
	\caption{Temperature scans of the remanence of several MRG films with $x = 0.5$, used to determine the compensation temperature.} 
	\label{MvsT}
\end{figure}

The densities of the films were obtained from the low angle X-ray reflectivity data. Theoretical densities are calculated from the atomic formulae reduced to exactly four atoms per formula unit, using the experimentally determined a- and c-parameters to determine the cell volume.  Calculating densities for Mn$_2$Ru$_{0.5}$Ga, for example, by assuming full occupancy or two vacancies per unit cell unit gives 7900~kg/m$^3$ or 7100~kg/m$^3$, respectively. The density obtained from fitting the XRR pattern was 7800~kg/m$^3$. This shows that the vacancy assumption underestimates the density by 9~\%, whereas the experimental density agrees with that calculated for full occupancy to within about 1~\%. A comparison of experimental and calculated densities of nine films for full occupancy is shown in Fig. \ref{Density}, where it is seen that the values agree to within 3~\%. The densities predicted for the formulae with 0.3 or 0.1 vacancies and 0.1 interstitial Mn atoms are marked by yellow crossed bars. We conclude that all the crystallographic sites in MRG really are close to fully occupied. Overall, the density increases with increasing Ru content as expected. The number of valence electrons $n_v$ per formula unit is included in Table \ref{tab_1}. 

Net magnetisation and hysteresis were measured by SQUID magnetometry. Compensation temperatures were deduced from the change of sign of the remanent magnetisation in zero applied field found from thermal scans, after saturating in 5~T at 400~K. All samples bar one exhibited a compensation point. Some of the thermal scans are presented in Fig. \ref{MvsT}. Curie temperatures were all above 400~K, the temperature limit of our SQUID measurements.

\begin{figure}[h]
	\centering\includegraphics[scale=0.24]{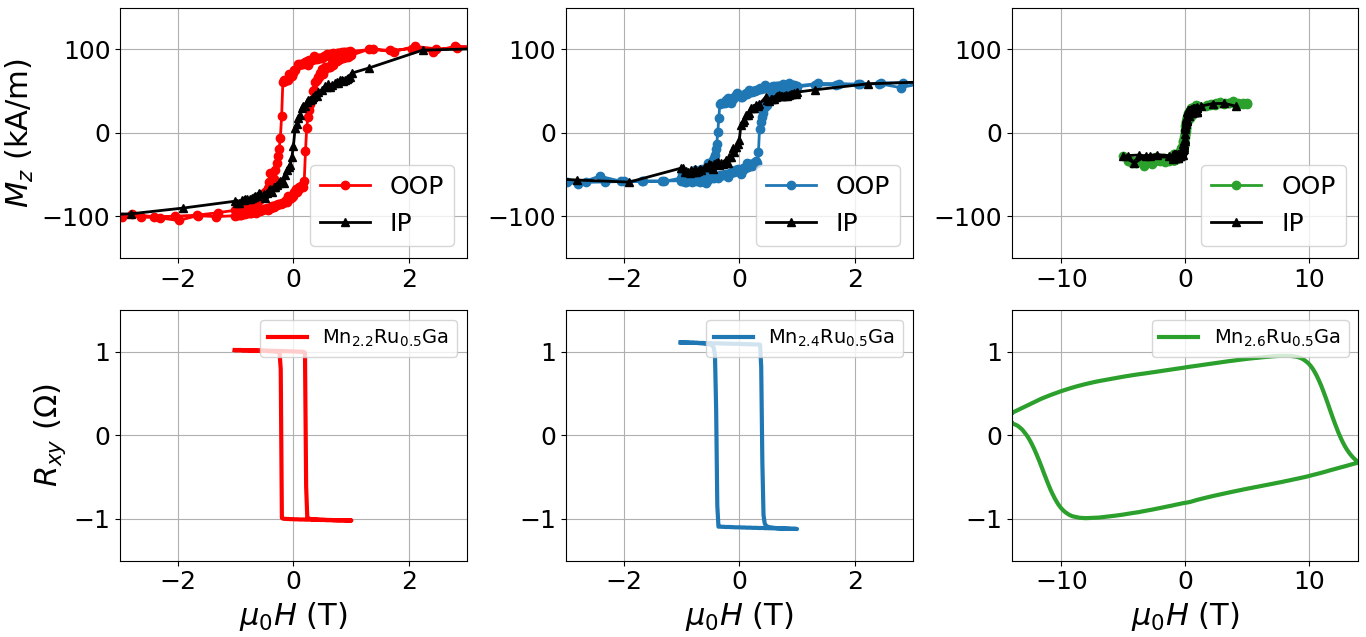}
	\caption{MRG thin films from for the $x = 0.5$ series. Top panel: Magnetisation and curves measured parallel (IP) and perpendicular (OOP) to the film plane. Bottom panel: Corresponding anomalous Hall effect curves. Note the applied field in the third panel on the right ranges from -14~T to +14~T.} 
	\label{MvsH_AHE}
\end{figure}
Hysteresis in these films can be extremely large close to compensation because the anisotropy field $\mu_0H_A = 2K_1/M_s$ diverges as the net magnetisation $M_s$ falls to zero. The net magnetisation measured by SQUID magnetometry with the field applied parallel or perpendicular to the film plane is shown for three $x = 0.5$ films in the top panel of Fig. \ref{MvsH_AHE}. The coercivity in the out of plane loop agrees with that seen in the anomalous Hall effect (Bottom panel). All in-plane loops exhibit a soft component that saturates in about 50~mT. This behaviour will be discussed in Section \ref{sec:level5}.
\begin{figure}[htbp]
	\centering
	\subfloat{\label{AHE_T}\includegraphics[scale=0.2]{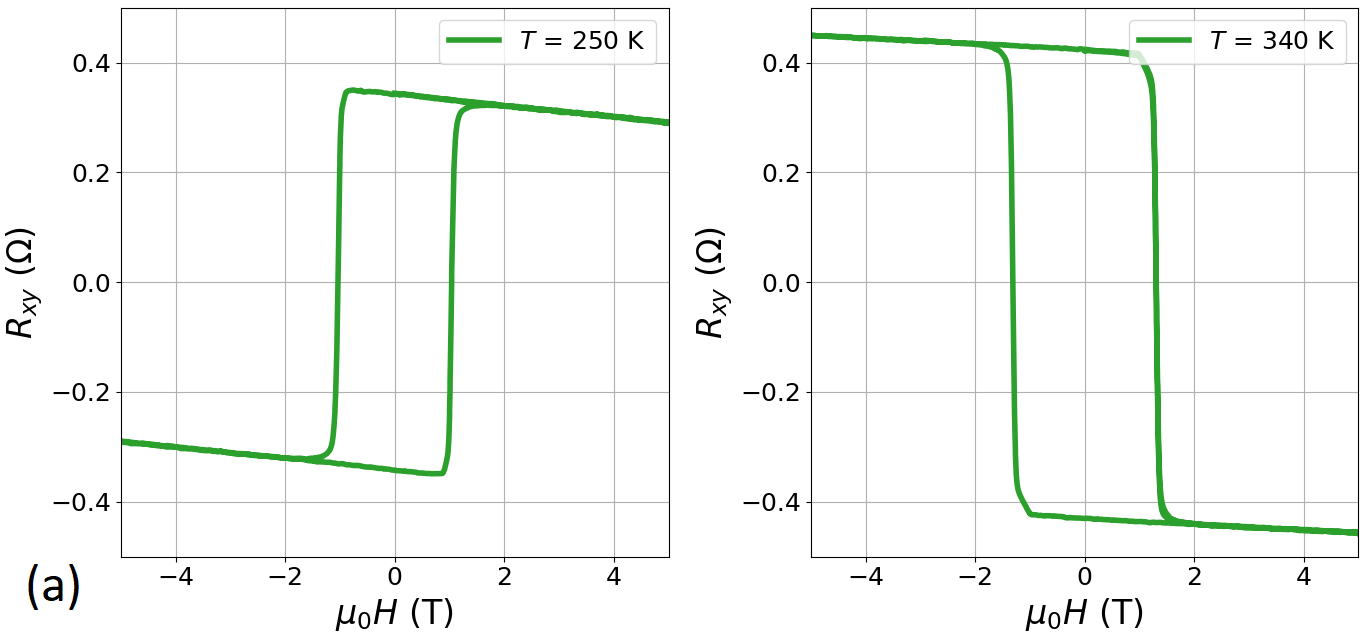}}\\
	\subfloat{\label{Angle}\includegraphics[scale=0.2]{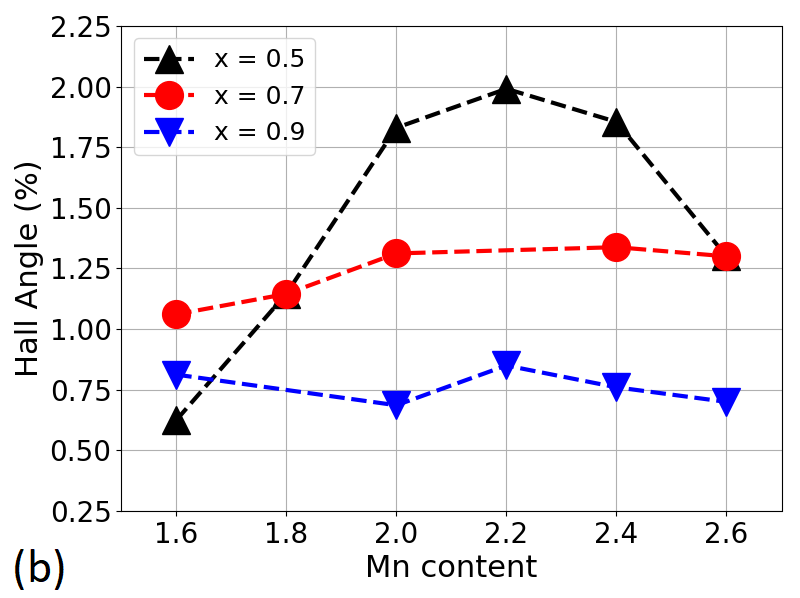}}
	\subfloat{\label{PCAR}\includegraphics[scale=0.2]{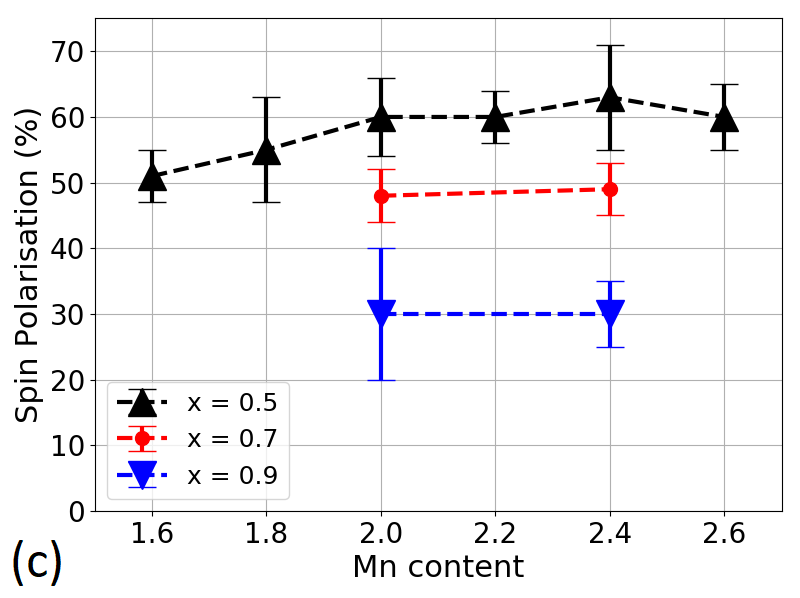}}\\
	\subfloat{\label{PA_x}\includegraphics[scale=0.2]{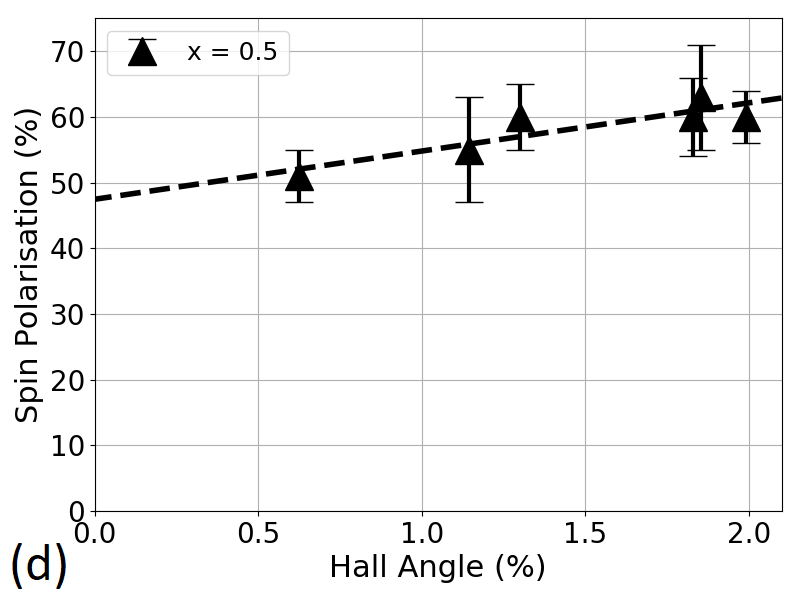}}
	\subfloat{\label{PA_y}\includegraphics[scale=0.2]{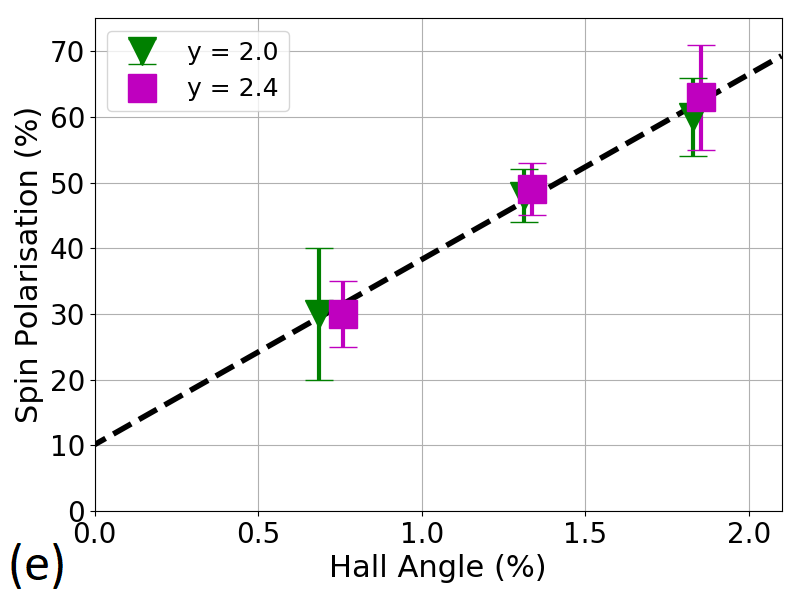}}\\
	\caption{Anomalous Hall effect: (a) anomalous Hall hysteresis loops for Mn$_{2.6}$Ru$_{0.5}$Ga below (250~K), and above (340~K) compensation (\tcmp~=~278~K). (b) Hall angles and (c) spin polarisation for three series in an extended range of Mn composition. The spin polarisation versus anomalous Hall angle for constant (d) $x=0.5$ and (e) $y=$~2.0, 2.4. } 
\end{figure}

The anomalous Hall effect cannot provide the magnitude of the sublattice magnetisation as it results from spin-dependent scattering and band structure effects \cite{Nagaosa.2010}, but the coercivity is determined accurately, and the dominant sublattice is obvious from the loop shape.  Fig. \ref{AHE_T} shows the reversal of the Hall loop on crossing compensation. The Hall signal is dominated by the Mn$^{4c}$ sublattice that contributes the majority of states close to the Fermi level. This sublattice is magnetically dominant below, but not above, \tcmp as can be inferred from the sign of the anomalous Hall coefficient. Sample Mn$_{2.6}$Ru$_{0.5}$Ga compensates at 278~K, where the coercivity diverges. The minor loop at room temperature already exhibits a coercivity of 12~T (Fig. \ref{MvsH_AHE}, bottom right panel). The anomalous Hall angle at room temperature, defined as the ratio of the Hall resistivity at saturation to the longitudinal resistivity reaches a maximum of 2~\% for the $x = 0.5$ series when $y = 2.2$, as shown in Fig. \ref{Angle}. Spin polarisation data determined by point contact Andreev reflection are summarised in Fig. \ref{PCAR}. The $x = 0.5$ series (in black) shows a maximum spin polarisation at the Fermi level of 63~\% for $y = 2.4$. Alloys with a higher ruthenium content have a considerably smaller spin polarisation. The spin polarisation is linearly dependent on the anomalous Hall angle $\theta_{AH}$ for $x=0.5$ and $y=$~2.0, 2.4 as shown in Figs. \ref{PA_x} and \ref{PA_y}, respectively. This demonstrates that the band structure accounts for $\theta_{AH}$, which is dominated by the lowest concentration of carriers and confirms that Mn$^{4c}$ sublattice electrons dominate the transport. A summary of the crystallographic and magnetic properties on the nine MRG films is provided in Table \ref{tab_1}. All films show compensation, with the exception of Mn$_{2.2}$Ru$_{0.5}$Ga.

\begin{figure}[h]
	\centering\includegraphics[scale=0.3]{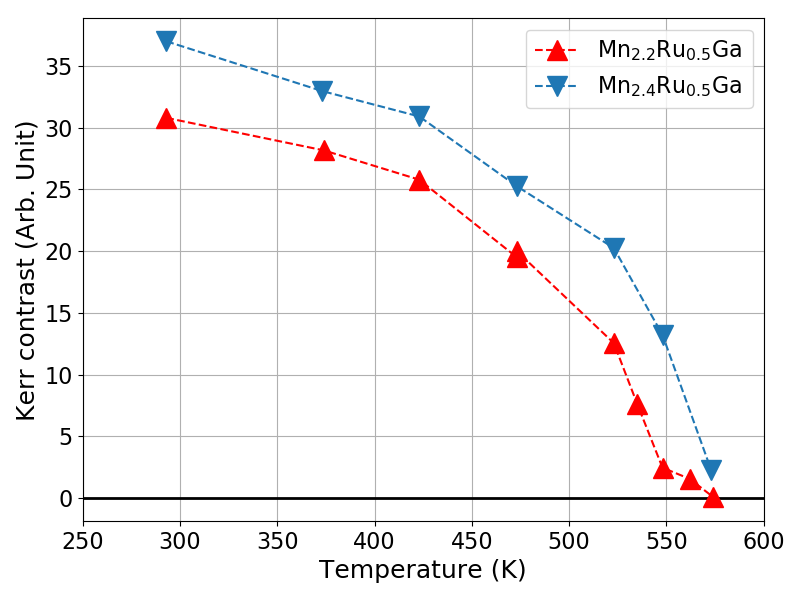}
	\caption{Temperature dependence of the optically dominant sublattice, measured by MOKE microscopy.} 
	\label{MOKE}
\end{figure}
MOKE signal shown in Fig. \ref{MOKE} is essentially that of the Mn$^{4c}$ sublattice magnetisation The MOKE contrast falls to zero at the ordering temperature, tabulated in Table \ref{tab_1}. The domain patterns in zero applied field tend to get frozen by defect networks, or whenever the hysteresis exceeds the stray field due to the small net magnetisation. \cite{Teichert2020}

\begin{table}[h]
	\caption{\label{tab_1} Compositions, number of valence electrons, crystallographic and magnetic properties of three series of MRG thin films where the Mn and Ru content are varied. $M_0$ and $m_0$ are magnetisation and moment at $T = 4$~K, respectively.}
	\begin{ruledtabular}
	\begin{tabular}{ C{1.0 cm}  C{0.6cm}  C{0.6cm}  C{0.6cm}  C{1.0cm}  C{0.9cm}  C{1.05cm}  C{0.7cm}  C{0.5cm} }
		Mn, Ru & $n_v$ & $c$ & $a$ & $\rho$ & $M_0$ & $m_0$ & $T_{\text{comp}}$ & $T_C$ \\
		 $y$, $x$ & & (pm) & (pm) & (kg/m$^{3}$) & (kA/m) & ($\mu_B$/f.u.) & (K) & (K) \\
		\hline 
		2.2, 0.5 & 24.2 & 605 & 597 & 8003 & -33 & -0.19 & -- & 550 \\
		2.4, 0.5 & 24.4 & 604 & 596 & 8135 & 6 & 0.03 & 130 & 577 \\
		2.6, 0.5 & 24.6 & 604 & 595 & 7840 & 38 & 0.22 & 278 & 592 \\
		2.0, 0.7 & 24.4 & 605 & 598 & 8296 & 13 & 0.08 & 165 & 530\\
		2.2, 0.7 & 24.6 & 605 & 596 & 8365 & 43 & 0.25 & 311 & 543\\
		2.4, 0.7 & 24.8 & 606 & 596 & 8359 & 67 & 0.39 & 381 & 494\\
		1.8, 0.9 & 24.6 & 608 & 596 & 8513 & 36 & 0.21 & 235 & 478\\
		2.0, 0.9 & 24.8 & 606 & 596 & 8409 & 61 & 0.35 & 375 & 513\\
		2.2, 0.9 & 25.0 & 607 & 597 & 8496 & 111 & 0.65 & 436 & 491\\
	\end{tabular}
\end{ruledtabular}
\end{table}

\section{\label{sec:level4}Mean Field Theory}

Zero-field thermal scans of net magnetisation measured by the SQUID obtained after saturating the magnetisation at 400~K were fitted using a two-sublattice Weiss molecular mean field model of colinear ferrimagnetism described by the system of non-linear equations
\begin{equation}
\begin{split}
	\textbf{H}^{4a} &= n_W^{aa}\textbf{M}^{4a} + n_W^{ac}\textbf{M}^{4c} \\
	\textbf{H}^{4c} &= n_W^{ac}\textbf{M}^{4a} + n_W^{cc}\textbf{M}^{4c} 
\end{split}
\end{equation}
where $H^{4a}$ and $H^{4c}$ are the internal sublattice molecular fields $M^{4a}$ and $M^{4c}$ are the magnetisations of the two sublattices, which are modelled by Brillouin functions and $n_W^{aa}$, $n_W^{ac}$ and $n_W^{cc}$ are the three Weiss coefficients. They are related to Heisenberg exchange constants $\mathscr{J}$ by the expression
\begin{equation}
	\mathscr{J}^{ij} = \frac{ n_W^{ij} \rho \mu_0 (g \mu_B)^2}{Z^{ij}}
\end{equation}
where $\rho$ is the number of the Mn atoms on the i$^\text{th}$ sublattice per unit cell volume, $\mu_0$ is permeability of free space, $g$ is the Land\'{e} g-factor for spin, $\mu_B$ is the Bohr magneton and $Z$ is the coordination number, where $Z^{aa} = Z^{cc} = 12$ and $Z^{ac} = 8$ for Mn$_2$Ru$_x$Ga. 

\begin{figure}[htbp]
	\centering
	\subfloat{\label{MFT_net}\includegraphics[scale=0.23]{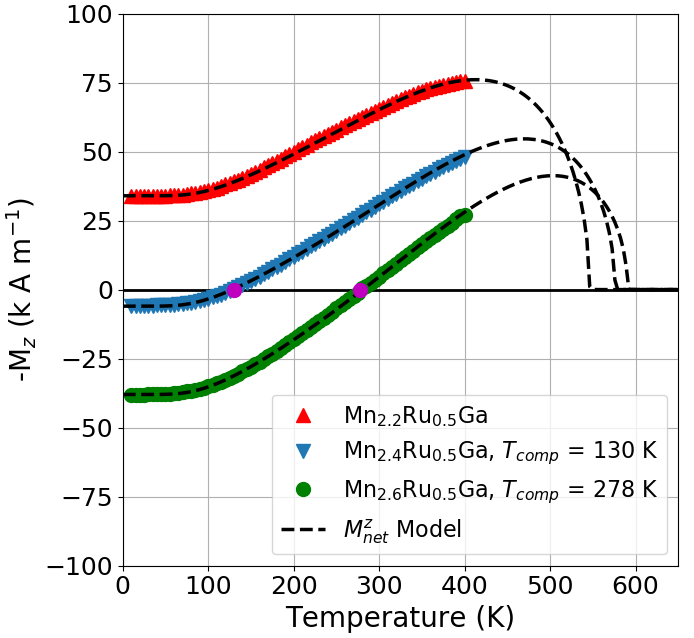}}
	\subfloat{\label{MFT_sub}\includegraphics[scale=0.23]{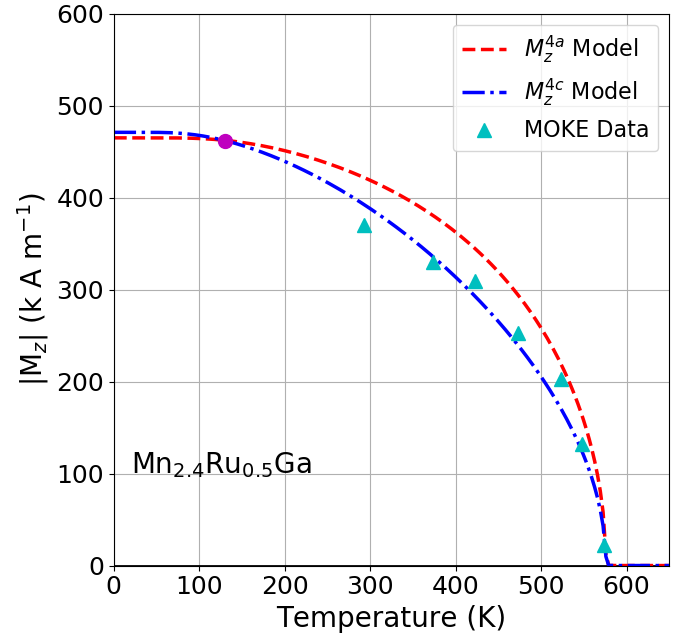}}
	\caption{(a) Mn$_{2.2}$Ru$_{0.5}$Ga, Mn$_{2.4}$Ru$_{0.5}$Ga and Mn$_{2.6}$Ru$_{0.5}$Ga thin films: Molecular field theory fit of net magnetisation versus temperature. (b) Mn$_{2.4}$Ru$_{0.5}$Ga: Calculated temperature dependences of Mn$^{4a}$ and Mn$^{4c}$ sublattice magnetisations compared with scaled MOKE data. The violet point marks compensation.} 
	\label{MFT}
\end{figure}

\begin{table}[h]
	\caption{\label{tab_2} Summary of sublattice magnetisation at $T = 0 $~K ($M_0$), spin angular momenta ($S^{4a,4c}$), Heisenberg exchange parameters from molecular field theory analysis of magnetisation data on the nine MRG thin films. The last column shows the ratio of the product of exchange constants with the coordination number $\mathscr{J}^{aa}Z^{aa}$ : $\mathscr{J}^{ac}Z^{ac}$ : $\mathscr{J}^{cc}Z^{cc}$.}
	\begin{ruledtabular}
	\begin{tabular}{C{1cm}  C{0.55cm}  C{0.55cm}  C{1.0cm}  C{0.7cm}  C{0.7cm}  C{0.7cm}  C{2.1cm}}
		Mn, Ru & $S^{4a}$ & $S^{4c}$ & $M_0^{4c}$  &$\mathscr{J}^{aa}$ & $\mathscr{J}^{ac}$ &$\mathscr{J}^{cc}$ & Ratio\\
		\hline
		$y$, $x$ &  &  & (kA/m) & (K) & (K) & (K) & \\
		2.2, 0.5 & 2.0 & 2.0 & 431 & 22.0 & -24.5 & 11.6 & 10.0 : -7.4 : 5.3\\
		2.4, 0.5 & 2.5 & 2.5 & 471 & 19.7 & -20.2 & 10.3 & 10.0 : -6.8 : 5.3\\
		2.6, 0.5 & 2.5 & 2.5 & 503 & 19.4 & -20.1 & 9.6 & 10.0 : -6.9 : 5.0\\
		2.0, 0.7 & 2.0 & 2.5 & 478 & 21.6 & -20.5 & 10.4 & 10.0 : -6.3 : 4.8\\
		2.2, 0.7 & 2.0 & 2.5 & 508 & 21.7 & -19.7 & 10.5 & 10.0 : -6.1 : 4.8\\
		2.4, 0.7 & 2.0 & 2.5 & 532 & 19.2 & -16.1 & 10.0 & 10.0 : -5.6 : 5.2\\
		1.8, 0.9 & 1.5 & 2.5 & 501 & 21.8 & -23.2 & 7.0 & 10.0 : -7.1 : 3.2\\
		2.0, 0.9 & 1.5 & 2.5 & 526 & 23.0 & -23.8 & 7.2 & 10.0 : -6.9 : 3.1\\
		2.2, 0.9 & 1.5 & 2.5 & 576 & 22.4 & -20.6 & 6.2 & 10.0 : -6.4 : 2.8\\
	\end{tabular}	
	\end{ruledtabular}
\end{table}

When more than half the sites per formula unit are manganese, we assume the excess populates the $4d$ sites which are symmetrically equivalent to the $4c$ sites. The $4a$ and $4b$ sites which accommodate Mn and Ga atoms are assumed to be fully occupied. The spin angular momentum quantum numbers for Mn atoms on each site, $S^{4a,4c}$, are multiples of $\frac{1}{2}$. The upper bound for $S^{4a,4c}$ is the sum of spins in a half-filled $3d$ orbital (5 spins). The lower bound is based on the Mn moment in similar compounds. (3 spins), so the quantum numbers in the Brillouin functions are chosen in the range $1.5 \leq S^{4a,4c} \leq 2.5$, and the sublattice magnetization is scaled to best represent the data. In metals, the spin moment per atom is rarely an integral number of Bohr magnetons and the sublattice moments at $T = 0$~K cannot be directly calculated from the $S^{4a,4c}$ quantum numbers appropriate for localised electrons due to orbital overlap and mixing. We found by density functional theory calculation that the Mn$^{4a}$ sublattice moment at $T = 0$~K is nearly independent of Ru content for $0.5 \leq x \leq 1.0$ with a value of $m_0^{4a} = 2.70~ \mu_B$/Mn, corresponding to a sublattice magnetisation $M_0^{4a} = 465$~kAm$^{-1}$.\cite{Zic.2016} The net magnetisation at $T = 0$~K was extrapolated from the SQUID magnetometry data, shown in Fig. \ref{MvsT}. Using this value, the magnetisation of the Mn$^{4c}$ sublattice was inferred.

MOKE microscopy at a fixed wavelength ($\lambda = \SI{632}{nm}$) measures the Kerr rotation, which for MRG is essentially proportional to the magnetisation of the Mn$^{4c}$ sublattice. Temperature-dependent MOKE data is compared with $M_z$ of the Mn$^{4c}$ sublattice calculated from the model in Fig. \ref{MFT_sub}, validating our approach. We use MOKE to determine $T_C$, since it covers the appropriate temperature range, and the Curie temperatures are then used as constraints in the fits.\cite{Smart1966} The SQUID data, MFT fits and MOKE data for $x = 0.5$ thin films are all shown in Figs. \ref{MFT_net} and \ref{MFT_sub}.

\section{\label{sec:level5}Discussion}

We define the Mn$^{4c}$ moment as positive, coupled magnetically to the negative Mn$^{4a}$. In a previous study it was shown that well below $T_C$ the Mn$^{4a}$ moment is almost temperature independent whereas the Mn$^{4c}$ moment varies nearly linearly as a function of temperature.\cite{Betto.2015} Therefore, adding (subtracting) atoms on the Mn$^{4c}$ sublattice is expected to raise (lower) \tcmp, whereas adding (subtracting) Mn atoms on the Mn$^{4a}$ will increase (decrease) the sublattice moment, but since this moment is negative it results in a decrease (increase) of \tcmp.

\begin{figure}[h]
	\centering
	\subfloat{\label{nv_tcomp}\includegraphics[scale=0.35]{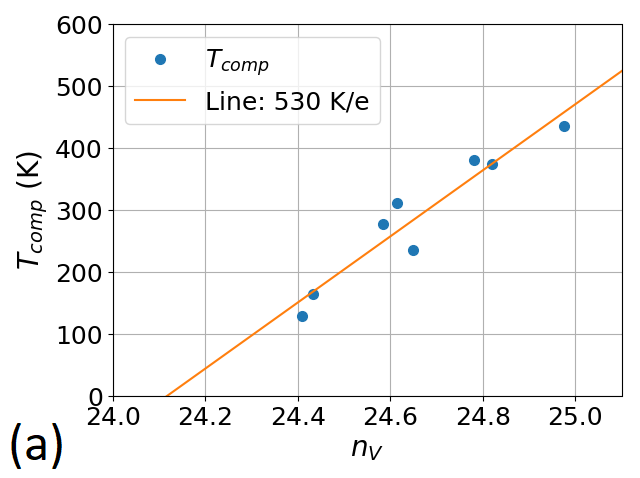}}\\
	\subfloat{\label{nv_m0}\includegraphics[scale=0.35]{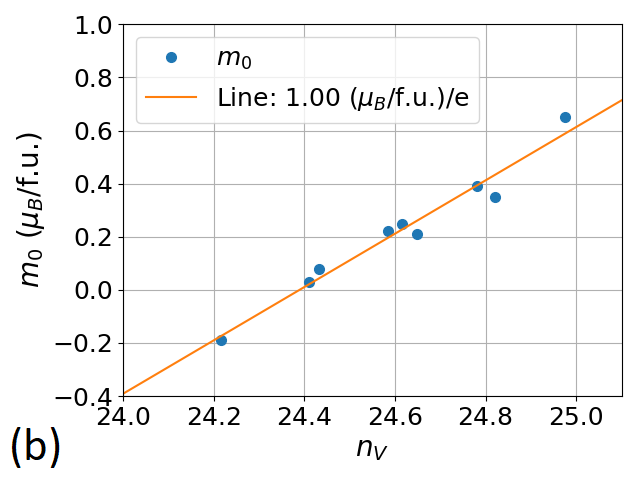}}
	\caption{(a) \tcmp~versus $n_V$: adding an electron ($e$) raises \tcmp~ by 530~K. (b) $m_0$ versus $n_v$: adding an electron ($e$) raises $m_0$ by $1\mu_B$/f.u. in agreement with the Slater-Pauling rule for a half-metal. } 
\end{figure}

Increasing Ru raises \tcmp~\cite{Kurt.2014} and within each Ru series increasing Mn also increases \tcmp. Fig. \ref{nv_tcomp} shows the relationship between \tcmp~and the number of valence electrons $(n_v)$, with a slope of 530~K per electron. We conclude that the added Mn predominantly increases the Mn concentration on the Mn$^{4c/4d}$ sublattice. This is supported by the analysis of the densities of the thin films plotted in Fig. \ref{Density}, which shows that there is no appreciable vacancy concentration for any composition in the series. Fig. \ref{nv_m0} shows the linear relationship between $m_0$ and $n_v$ with a slope of $1\mu_B$/f.u. per electron. 
The Slater-Pauling behaviour confirms the half-metallicity of all the MRG thin films. 

The magnetometry data in Fig. \ref{MvsT}, was recorded after magnetically saturating the samples perpendicular to their surface at $T = 400$~K. The $z$-projection of the net magnetisation was measured in zero field as a function of temperature during cool down. An ideal data set would include all three components of the sublattice specific moments which might be obtained using neutron diffraction or, if two components would suffice, by x-ray magnetic circular dichroism (XMCD). \cite{Siewierska2021} Here we first discuss the magnetic mode of MRG first in the simplified model of two collinear sublattices, based on known variations of the Mn-Mn exchange parameters with the distance between interacting atoms.

Table \ref{tab_2} summarises the fit parameters for the series. For all compositions, the model agrees well with the magnetisation data measured along the easy axis, perpendicular to the film plane. The ratios of Heisenberg exchange constants are similar to those was reported previously for an MRG film with a comparable composition. \cite{Davies2020} The main trend that emerges from the data is that as the Ru content of MRG increases, $\mathcal{J}^{cc}$ decreases and while $\mathcal{J}^{aa}$ and $\mathcal{J}^{ac}$ remain almost constant. We explain the results as follows:  $\mathcal{J}^{aa}$ does not depend on Ru content because Mn$^{4a}$ does not have much overlap with Ru$^{4d}$. When the number of Mn atoms in the unit cell exceeds 16 some Mn fills the $4d$ positions, giving rise to Mn$^{4c}$ -  Mn$^{4d}$ interactions, thereby the weakening $\mathcal{J}^{cc}$. 
\begin{figure}[htbp]
	\centering
	\subfloat{\label{Ball_order}\includegraphics[scale=0.21]{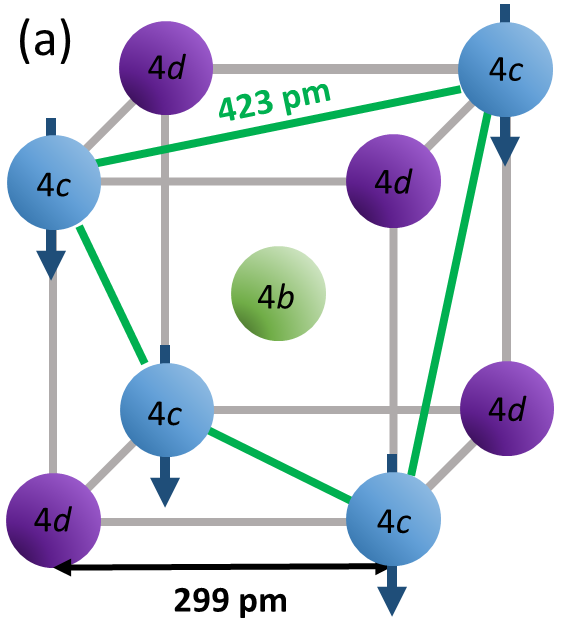}}
	\subfloat{\label{Ball_disorder}\includegraphics[scale=0.21]{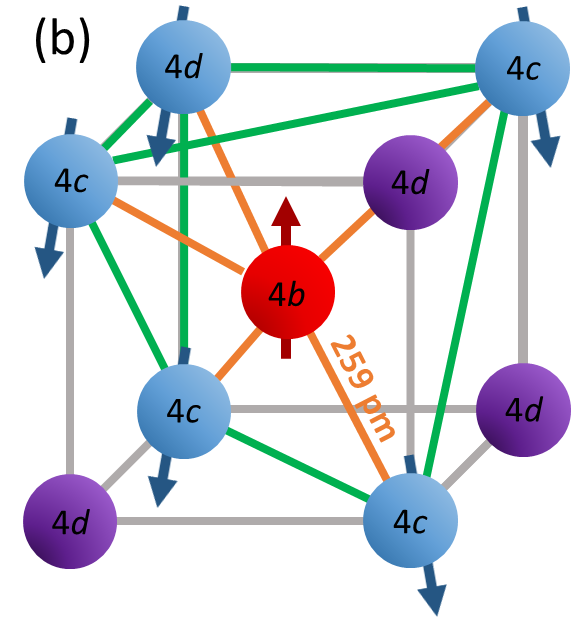}}	
	\subfloat{\label{arrows_cone}\includegraphics[scale=0.182]{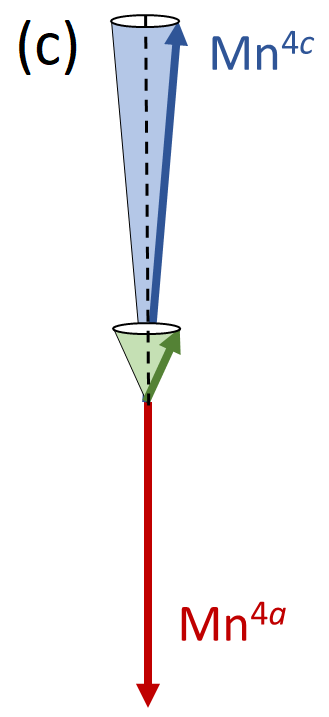}}
	\caption{Left: Inner cube (Mn$^{4c}$ sublattice) in the cubic unit cell of the inverse Heusler structure \textit{XA}  (Fig. \ref{figure:model}) showing the (a) $4c$ and $4d$ sites occupied by Mn atoms (blue with arrow) and Ru atoms (purple) with Ga (green) in the center, (b) one Ru atom is replaced with an Mn atom on a $4d$ site and Ga is replaced with Mn on the $4b$ site (red with arrow). Nearest neighbour ferromagnetic and antiferromagnetic couplings are represented by green and orange lines, respectively. The competition between exchange couplings can induce non-collinearity of sublattice moments locally. (c) Illustration of canting of net moment due to Mn$^{4c}$ sublattice non-collinearity.} 
	\label{Model}
\end{figure}

It is known that Mn-Mn distances in the range 250~pm - 280~pm generally lead to antiferromagnetic coupling that decreases in magnitude as the distance increases, whereas at distances greater than 290~pm the coupling becomes ferromagnetic. \cite{Coey_book} In an unstrained Mn$_2$RuGa film, $a_{0} =  598$~pm. The intra-sublattice distances between Mn atoms, where each site forms an fcc sublattice, are $a_0/\sqrt{2}\approx 423 $~pm and the inter-sublattice Mn$^{4a}$ - Mn$^{4c}$ or Mn$^{4b}$ - Mn$^{4c}$ distance is $a_0 \sqrt{3}/4 \approx 259 $~pm, as indicated in Fig. \ref{Model}. These distances correspond to ferromagnetic and antiferromagnetic exchange coupling, respectively. They are changed a little by the tetragonal distortion Since the densities indicate that all sites are occupied, chemical disorder will arise in the unit cell. A first type of disorder arises from Mn$^{4a}$-Ga$^{4b}$ antisites and a second is the presence of Mn atoms on $4d$ sites. The Mn$^{4c}$ sublattice portion of the unit cell is shown for Mn$_2$RuGa and Mn$_{2.25}$Ru$_{0.75}$Ga in Figs. \ref{Ball_order} and \ref{Ball_disorder}, respectively, where the latter illustrates both types of disorder. The presence of Mn on $4d$ sites gives rise to FM coupling, marked with green lines in Fig. \ref{Ball_disorder}. Changing the positions of Ga$^{4b}$ and Mn$^{4a}$ results in AFM coupling because the Mn$^{4b}$ - Mn$^{4c}$ distance is $\approx 259 $~pm. This leads to random competing exchange on the Mn$^{4c}$ sublattice, which will tend to give rise to a non-collinear spin structure. The local Mn$^{4c}$ macroscopic sublattice moment may be expected to cant away from the anisotropy axis, forming an easy cone if four-fold in-plane anisotropy is negligible. The cone angle is the angle between the anisotropy axis and the moment. Local atomic environments differ on account of the random site occupancies that follow from the composition of the films, adding an element of randomness to the competing interactions, and the local canting angle.
 
In-plane applied field SQUID magnetometry data in Fig. \ref{MvsH_AHE}, shows a component of the net moment which easily saturates along the \textit{in-plane} field direction, unlike the hard component which normally extrapolates to the anisotropy field. For films with higher Ru content, which are closer to compensation at room temperature, the soft component dominates, and the determination of the anisotropy field is problematic. From the data on the first two samples in Fig. \ref{MvsH_AHE} we deduce a cone angle for the net moment of \SI{21}{\degree}.  We have measured the cone angle in other MRG thin films by vector SQUID magnetometry and find values that can be as large as \SI{40}{\degree}. A feature of the \textit{out-of-plane} loops in Fig. \ref{MvsH_AHE} is the step in magnetisation near remanence that corresponds to 40~\% of the weak moment. We associate this with a closing of the Mn$^{4c}$ cone and a simultaneous opening of a cone on the Mn$^{4a}$ sublattice, for which we have seen evidence in XMCD data. \cite{Siewierska2021}

From the analysis of exchange energies and lattice parameters, it follows that sublattice non-collinearity results from competing positive and negative exchange coupling for the Mn$^{4c}$ sublattice. This case has previously been argued for Mn$^{2d}$ sites in tetragonal D0$_{22}$ Mn$_2$RhSn.\cite{Meshcheriakova.2014} A picture that represents our experimental observations well appears on the right hand side of Fig. \ref{arrows_cone}. We note that the Mn$^{4c}$ cone angle needs to be only a few degrees to result in canting of the net moment by tens of degrees, because the net moment is more than an order of magnitude smaller than the sublattice moments, depending on how close the temperature is to compensation. The AHE loops in Fig. \ref{MvsH_AHE}, where $R_{xy}$ is proportional to the perpendicular component of the Mn$^{4c}$ sublattice magnetisation, exhibit high remanence and we conclude that the very small in-plane moment of the Mn$^{4c}$ sublattice cannot be detected. 

\begin{figure}[h]
	\centering\includegraphics[scale=0.25]{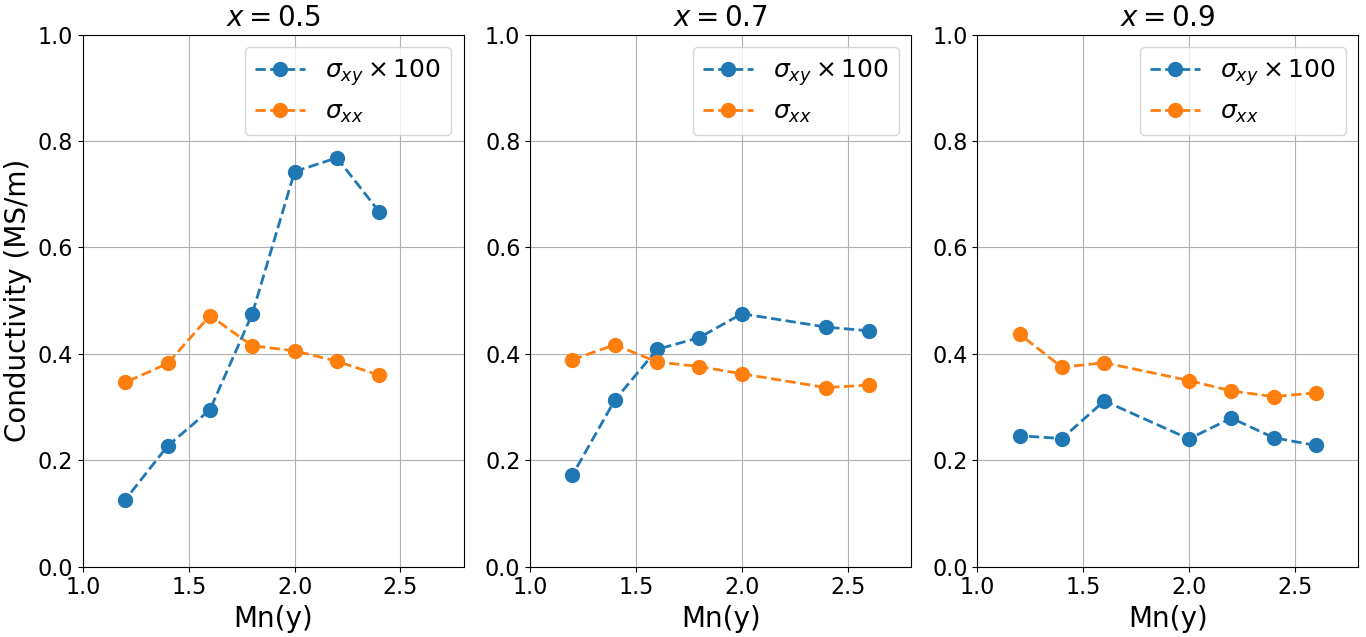}
	\caption{Transverse ($\sigma_{xy}$) and longitudinal ($\sigma_{xx}$) conductivities of MRG thin films at 300~K for three Ru($x$) series versus Mn($y$) content. Note that $\sigma_{xy}$ was calculated at magnetic saturation in $\mu_0 H = 1$~T at $T = 300$~K and in the data in the graph has been multiplied by a factor of 100.} 
	\label{Hall_Cond}
\end{figure}

The longitudinal ($\sigma_{xx}$) and transverse ($\sigma_{xy}$) conductivities of MRG thin films are plotted for three Ru($x$) series versus Mn($y$) in Fig. \ref{Hall_Cond}. The magnitude of $\sigma_{xx}$ in the absence of a magnetic field is proportional to the conduction electron scattering. In all samples, $\sigma_{xx}$ is almost constant whereas $\sigma_{xy}$ shows a strong dependence on both Mn and Ru. To explain the observations, we need to consider the origins of AHE. There are two types of contributions to the anomalous conductivity. The first is the extrinsic contribution which is scattering dependent. There are two distinct mechanisms, skew scattering and side jump, where the former is proportional to $\sigma_{xx}$ and the latter is independent of $\sigma_{xx}$. The second contribution is scattering independent (intrinsic) arising from Berry curvature and is independent of $\sigma_{xx}$. \cite{Nagaosa.2010}

In Fig. \ref{Hall_Cond}, $\sigma_{xy}$ exhibits a decrease with increasing Ru content. When Mn is varied with low $x$, $\sigma_{xy}$ is seen to be independent of $\sigma_{xx}$, but for $x = 0.9$ $\sigma_{xy}$ is approximately constant. The independence of $\sigma_{xy}$ on $\sigma_{xx}$ and its strong dependence on Ru content $x$ suggests that for low $x$ films the main contribution to AHE has an intrinsic origin. Impurity scattering, which depends on spin-orbit interaction, would be expected to increase with $x$. We find a maximal anomalous Hall angle $\theta_{AH} = 2$~\% for Mn$_{2.2}$Ru$_{0.5}$Ga, which is $\approx 10$ times higher than in conventional ferromagnets measured at 300~K. \cite{Seki2019} Such large Hall angles are consistent with a large intrinsic contribution. While this could be a Berry phase effect, we are unable to disentangle the possible contributions of band structure (Weyl points), non-collinearity, Fermi surface effects and the influence of the incipient spin flop, which is expected for a ferrimagnet near compensation \cite{Ting_2021}, and which is complete in Mn$_{2.2}$Ru$_{0.7}$Ga at 7.65 T \cite{Siewierska2021}. \\

\section{\label{sec:level6}Conclusions} 

At the outset Mn$_2$Ru$_{0.5}$Ga was assumed to have vacancies on the $4d$ sites. Here we have shown from the observed densities that all four sites in MRG are all practically full, which implies that Mn$_{2}$Ru$_{0.5}$Ga has 24 valence electrons, not 21 as originally thought. Full occupancy of the  $4a$ and $4c$ sites by Mn leads to a collinear ferrimagnetic structure, which accounts for the magnetic material properties. 

The substantial absence of vacancies implies the presence of chemical disorder. Specifically, the presence of some Mn on  $4b$ and $4d$ sites promotes antiferromagnetic intra-sublattice exchange coupling on the Mn$^{4c}$ sublattice, which leads to a non-collinear ferrimagnetic structure. The non-collinearity or the the small net moment is much more pronounced  than of the Mn$^{4c}$ sublattice moment. The discrepancy between the AHE and measurements of magnetisation in Fig. \ref{MvsH_AHE} can therefore be explained. The independence of longitudinal and transverse conductivities indicates the dominance of the intrinsic contribution to AHE, which accounts for the large anomalous Hall angle observed in MRG films with low $x$ . The reduction in spin polarisation with increasing Ru corresponds to a narrowing of the spin gap in the density of states. The half-metallicity of MRG in the range of compositions investigated is confirmed by Slater-Pauling behaviour of the net moment. 

The magnetisation, spin polarisation and compensation point of MRG can be tuned to match the requirements of a specific application by varying the composition of the films. Highly crystalline, textured thin films with magnetic compensation ranging from $T = 0$~K up to the magnetic ordering temperature can be produced. The coercivity near compensation that can exceed 10~T, could permit the incorporation of single MRG layers in thin film stacks without any additional antiferromagnetic layers. \cite{Borisov.2016}

\begin{acknowledgments}
K.E.S. and J.M.D.C. acknowledge funding through Irish Research Council under Grant GOIPG /2016/308 and in part by the Science Foundation Ireland under Grants 12/RC/2278 and 16/IA/4534. G.A., A.J. R.S. S.L, J. O'B., P.S. and K.R. acknowledge funding from TRANSPIRE FET Open Programme, H2020. K.E. acknowledges funding from Science Foundation Ireland US-Ireland R\&D (Center-to-Center) under grant 16/USC2C/3287. N.T. was supported by the European Union's Horizon 2020 Research and Innovation Programme under Marie Sk\l{}odowska-Curie EDGE Grant 713567, and in part by Science Foundation Ireland under grant 16/IA/4534.
\end{acknowledgments}

\bibliography{Refs_MRG}

\end{document}